\documentclass[11pt]{article}
\usepackage{authblk}
\usepackage{geometry}                
\usepackage{graphicx}
\usepackage{amssymb,extarrows}
\usepackage{epstopdf}
\usepackage[inline]{enumitem}

\newcommand{\pr}{\text{Pr}}

\newcommand{\be}{\begin{equation}}
\newcommand{\ee}{\end{equation}}
\newcommand{\bes}{\begin{equation*}}
\newcommand{\ees}{\end{equation*}}

\title{Open Problems in Mathematical Biology}

\author{Sean T. Vittadello$^{1,2}$ \& Michael P.H. Stumpf$^{1,2,3}$}
\affil{$^1$Melbourne Integrative Genomics, University of Melbourne, Australia\\ $^2$School of BioSciences, University of Melbourne,  Australia\\
$^3$School of Mathematics and Statistics, University of Melbourne, Australia}
\begin{document}
\maketitle
\begin{abstract}
Biology is data-rich, and it is equally  rich in concepts and hypotheses. Part of trying to understand biological processes and systems is therefore to confront our ideas and hypotheses with data using statistical methods to determine the extent to which our hypotheses agree with reality. But doing so in a systematic way is becoming increasingly challenging as our hypotheses become more detailed, and our data becomes more complex. Mathematical methods are therefore gaining in importance across the life- and biomedical sciences. Mathematical models allow us to test our understanding, make testable predictions about future behaviour, and gain insights into how we can control the behaviour of biological systems. It has been argued that mathematical methods can be of great benefit to biologists to make sense of data. But mathematics and mathematicians are set to benefit equally from considering the often bewildering complexity inherent to living systems. Here we present a small selection of open problems and challenges in mathematical biology. We have chosen these open problems because they are of both biological and mathematical interest.  
\end{abstract}
\section{Introduction}
In 1202 Leonardo Pisano (aka Fibonacci) studied the population dynamics of rabbits in his book {\em Liber Abaci}. By some, admittedly glib, measure mathematical biology could be said to predate theoretical physics --- which also started in Pisa with the work of Galileo Galilei --- by some 400 years. Arguably this head start has been lost and for the past century physics and mathematics have become deeply entwined. Biology, by contrast, was often dominated by observational or reductionist approaches and mathematical methods have for a long time only played a negligible role. Biologists have, however, been quick to adopt quantitative methods and tools as soon as there was a clear use for them. Biological problems and analyses have, for example, been important drivers of statistical methodology development, and statistics has become a main feature --- often not well liked --- of biology undergraduate and graduate education. Mathematics, less so. 
\par
The reasons why mathematics is a useful tool for many (not all) biological questions have been superbly laid out before \cite{May:2004p11952,cohen2004,Bizzarri:2019aa}. The principal reason, arguably, is that through the lens of mathematics we can make accessible the structures and mechanisms underlying heterogeneous and often disparate data and observations. If, for example, we can connect two observations or measurements in one model (say the level of growth factor in a Petri dish or well containing cells and the down-stream change in phosphorylation of nuclear ERK \cite{Cursons:2015wf,Filippi:2016gs}) we have a non-trivial mechanistic hypothesis from which we initiate further investigations. The analogy between mathematics and a microscope \cite{cohen2004} in making the unobserved or unobservable visible holds surprisingly often. 
\par
Our focus here is therefore different: we are interested in what are the fundamental challenges for mathematical biology right now. Any such discussion will be subjective and incomplete. In our case it is clearly also influenced by our respective backgrounds (pure mathematics and theoretical/statistical physics), our past and current research interests (which are pretty broad, but mostly concerned with cellular processes and biophysics), and the conversations that we have had with researchers over the years. One key person was Edmund Crampin, especially for MPHS. The selection of material here cannot claim completeness, but we do think that all the problems touched upon here would have been of interest to Edmund. We hope that this will also be of wider interest.  
\par
Our discussion below is broken into four parts: (i) A brief discussion as to the different demands on modelling and the modeller in biology and physics; (ii) A selection of open problems in mathematical biology. These questions were partly crowd-sourced, and we attempt to put them into the wider context of mathematical modelling in biology and the biomedical sciences. This part contains open problems that are concerned with making mathematical modelling more useful and relevant to biological problems, and open problems where new mathematics will be essential to find answers to long-standing open questions at the core of biological research. We then (iii) provide an, arguably opinionated, overview of the opportunities that exist for ideas and concepts primarily from pure mathematics, but also theoretical physics to enrich the toolbox of mathematical biologists and study aspects that are particular to living systems.

\section{Why is Modelling Biology so Difficult?}
The question really refers to a comparison with physics (and parts of chemistry): why is modelling biology apparently so much more difficult than modelling physics? There are three aspects to our attempt at an answer: (i) physicists can often rely on first principles to model systems of interest; (ii) in biology we often have to start from heuristic models; and finally, (iii) differences in education between physics and biology. 
\par
Our attempts at finding an answer to this question should start by pointing out that not all of physics is easy to model. For many physical problems, however, it is possible to write down the structure of a model, often starting from first principles \cite{Thorne:2017tm,Weyl:1952uz}. While it took major intellectual feats to write down, for example, Kepler's laws, Newton's laws, Maxwell's equations, Einstein's theories of relativity, and the laws of quantum mechanics and quantum field theory, the structure of a model is typically derivable from first principles. Noether's Theorem \cite{Goldstein:2002up} states that every conservation law has an associated symmetry: for example, conservation of energy is related to invariance with respect to time. Invariance with respect to position, or translational invariance, implies the conservation of momentum. In both a real and an abstract sense, Noether's theorem underlies much of what is often perceived as the beauty of physics; and being able to rely on it (the Ward--Takahashi identities generalise the classical picture into the domain of quantum field theory) makes the life of the working physicist much easier. 
\par
For dissipative systems, that is for systems where energy is consumed, Noether's theorem does not apply in its simple form. Biology is replete with systems that consume and turn over energy, a fundamental aspect of life. This fact alone suffices to account for the profound difficulties that arise in finding mathematical descriptions of biological systems and processes. 
\par
In practical terms the main difference between modelling physics and modelling biology is that the mathematical form of the model is known in most physical cases. It may be difficult to solve the many particle Dirac Equation, or the Navier--Stokes equation in 3D for realistic viscosity, but the form or type of equations is rarely uncertain. If it is, however, for example systems with broken symmetry \cite{PWAnderson1972}, then it is typically of utmost physical interest to identify and impose the correct form of the model. 
\par
In biology, by contrast, we have to start from a different level of abstraction, and the best model structure is rarely if ever known with any degree of certainty. We can make pragmatic assumptions to develop some simplified models, and the canonical model of this approach is clearly the Lotka--Volterra model
\begin{align*}
\frac{dx}{dt} &= \alpha x -\beta x y,\\[0.1cm]
\frac{dy}{dt} &= \gamma x y -\delta y
\end{align*}
where $x$ and $y$ denote a prey and a predator species respectively\cite{May:2004p11952}. The rate parameters, $\alpha$, $\beta$, $\gamma$, and $\delta$ represent, in turn, the prey growth rate, the decline in the prey population due to predation, the growth of the predator population, and the reduction in the predator population due to predator death. This model has been successful in explaining many aspects of population biology and ecology. But it is highly abstracted and idealised, and fundamental features of reproduction, predation, and so forth are purposefully ignored in this simple model. But this simple Lotka--Volterra model can form the basis for more complicated descriptions, including spatially structured populations, overlapping niches, and multiple trophic levels. We can use these simple models to obtain qualitative insights into fundamental aspects of biological processes and phenomena, including bet-hedging and the evolution of virulence. The Lotka--Volterra model is sufficiently easy to allow mathematical analysis, and simple enough to be accessible to people with little direct modelling experience. Progress in physics has often relied on a small canon of models that are equally relevant and accessible to theoreticians and experimentalists: the single harmonic oscillator, Ising Model, particle in a box, H$_2$ molecule, etc. Biology needs more of these \cite{MacArthur:2022ww}.
\par
Finally, the training tradition in physics is different: the typical physics curriculum is heavy on mathematics and mathematical methods (with considerable differences between institutions and even nations) that prepare students for theoretical physics. Over the past generation the mathematics curricula of physics degrees appear to have changed comparatively little in substance, but all physicists are exposed to the mathematical techniques that are necessary to make progress in research. The situation in biology is very different: life-scientists often receive training in qualitative and statistical methods required to make sense of observational and experimental data, whereas there is generally no consideration of modelling the behaviour of biological systems. The mathematical tools required by mathematical biology, meanwhile, are still proliferating; the diversity of problems considered and the complexity of data being generated pose considerable technical and conceptual challenges which require a broad mathematical tool-set. We pick up this point in Section~\ref{sec:new}.
\par


\section{Open Problems in Mathematical Biology: A Small Selection.}\label{sec:open}
\subsection{A Note on Notation}
We are considering models, $f(x;\theta)$, describing the state $x\in\Omega\subseteq \mathbb{R}^N$ of a biological system; the states $x$ are confined to the state space, $\Omega$, of the system.  Here $\theta\in \Theta \subseteq \mathbb{R}^d$ is the $d$-dimensional vector of model parameters (e.g. reaction rate parameters) in the parameter space $\Theta$. We denote by $x_i$ and $\theta_i$ the $i$th components of the state and parameter vectors, respectively.
 \par  
When $f(x;\theta)$ represents a dynamical system, for example an ordinary differential equation $dx/dt = f(x;\theta)$, we use $x_0\in \Omega$ to denote the initial conditions. Alternative models will be represented by subscripts, $j$, therefore $f_j(x;\theta)$.
\par
We represent any experimental observations or data by $\mathcal{D}=\{d_1,d_2,\ldots,d_m\}$; here $d_i$ is a vector with number of coordinates possibly less than $N$ to account for observations that do not capture the whole state space, therefore where $\dim(d_i)<\dim(x)$.

\subsection{Inverse Problems and Parameter Sensitivity}
In many situations we have a candidate model, $f(x;\theta)$, or at least its structure. What we then require are the parameters, $\theta$, and on occasion initial conditions, $x_0 = x(0)$. Choice of parameters is crucial if we want to compare model output with observed behaviour or data $\mathcal{D}$. Often parameters are chosen from the literature, but this approach is fraught with potential problems \cite{Kirk:2015gj}. Literature information can be imprecise about the conditions under which parameters were determined; if, for example, ambient pH is not comparable then the reaction rate may differ from the conditions under consideration. Combining reaction rate parameters from different sources is problematic if the corresponding experimental conditions were not identical or at least comparable. So in many instances we will have to infer parameters from data \cite{Villaverde:2019ws,Villaverde:2016vr,Gabor:2015wd,Liepe:2014iwa,Toni:2009tr}.
\par
We typically have three options to delimit the range for the parameter $\theta$. We can identify the value of $\theta$ that produces model output that is most similar to the observed data. The likelihood,
\begin{equation}
L(\theta) = \pr(\mathcal{D}|\theta),
\end{equation}
is the probability of the data for a given parameter $\theta$. The maximum-likelihood estimate (MLE), $\hat{\theta} = \text{argmax}_\theta(L(\theta))$, corresponds to the parameter that has the highest probability of producing the observed data, and it can be obtained through optimisation. We usually use optimisation to estimate 
$\hat{\theta}$. In addition to this point-estimate we can determine confidence intervals for the value of $\theta$. 
\par 
In a Bayesian framework we determine the posterior probability of a parameter,
\be
\pr(\theta|\mathcal{D}) = \frac{\pr(\mathcal{D}|\theta)\pi(\theta)}{\pr(\mathcal{D})},
\ee
where the prior, $\pi(\theta)$, represents our knowledge (or assumptions) about the parameter value prior to looking at any data. The posterior is thus determined from both prior information and the available data, $\mathcal{D}$, via the likelihood, $\pr(\mathcal{D}|\theta)$. The probability distribution over parameters rather than a single estimate tends to be the focus of Bayesian inference. Both frequentist and Bayesian inference are vast areas of research in their own right.
\par
What really drives inference, irrespective of the framework, is the information content of the data (assuming that the model structure, that is the functional form of $f(x;\theta)$, is sufficiently close to the truth). One measure of this is how much varying $\theta$ changes the probability of observing the data, therefore the likelihood or the posterior probability. Information geometry arguably provides the most rigorous way of assessing certainty in an estimate, especially the so called Fisher information \cite{Komorowski:2011cn}. Under mild regularity conditions the Fisher information matrix is given by
\be
[I(\theta)]_{i,j} = - E\left[\frac{\partial^2}{\partial \theta_i\partial \theta_j} \log L(\theta)\left|\theta\right.\right].
\ee 
The curvature of the the log-likelihood, or posterior function (or the curvature of a different cost function), is an appropriate measure of certainty for uni-modal likelihoods. For multi-modal likelihoods/posteriors {\em ad-hoc} measures have been proposed, but no easy or universally satisfactory solution exists \cite{Secrier:2009tu,Daniels:2008vr}. 
\par
What we do find in many practical applications is that for any given dataset and model we can only infer some of the parameters (more generally some combinations of parameters); the rest are simply not inferable with any precision, for example their marginal posteriors do not differ substantially from the posteriors. Identifying which parameter combinations can be learned \cite{Liepe:2013vj} from prior knowledge plus data will become a crucial part of parameterising large models, including whole cell models \cite{Babtie:2017ix}. In the short term sensitivity and robustness analysis --- therefore varying parameters and assessing the resulting differences in model output --- can be used to assess which parameters can be inferred from data \cite{Nam:2020td}.

\subsection{Model Selection}
Parameter estimation is already a formidable problem. But generally, as we have shown above and will argue further below, the structure and nature of the correct model are unknown. Choosing which model $M_i$ from a set of models, $\mathcal{M}=\{M_1,\ldots,M_N\}$, does the best job of explaining data has therefore become a central focus in some areas of mathematical biology. A host of methods are available to enforce something akin to Ockham's razor: generating a model that is as complex as required but not more so. These methods have been reviewed extensively elsewhere \cite{Kirk:2013hq}. When parameters can be estimated from the data using likelihood or Bayesian methods, information criteria such as the Akaike information criterion (AIC) and the Bayesian information criterion (BIC) are popular because they explicitly penalise against models with more parameters:
\begin{align}
\text{AIC}_i &= -2\log L(\theta_i;M_i)+2k_i,\\[0.1cm]
\text{BIC}_i &= -2\log L(\theta_i;M_i)+k_i \log n,
\end{align}
where $k_i$ is the number of parameters of model $i$ and $n$ is the size of the dataset. They are both, however, approximations (and in the case of the AIC biased towards overly complex models as the amount of data increases) to the marginal likelihood of a model,
\be
\pr(\mathcal{D}|M_i) = \int_\Omega  \pr(\theta|M_i) \pr(\mathcal{D}|\theta,M_i) d\theta, 
\label{eq:marg}
\ee 
and the model posterior probability,
\be
\pr(M_i|\mathcal{D}) =\frac{\pr(\mathcal{D}|M_i)\pi(M_i)}{\pr(\mathcal{D})}.
\ee
Here the number of model parameters is {\em implicitly} penalised. 
\par
Model selection is challenging because of the high-dimensional integration problem in Equation~\eqref{eq:marg}. Any advances in constraining and guiding parameter inference --- such as by identifying subspaces of the full parameter space $\Theta$ containing parameters that do not affect the model fit to data --- will potentially help, but these subspaces will differ between alternative models \cite{Barnes:2011hh}. Specifying the prior probability of models is notoriously difficult: in principle there is no clear {\em a priori} bound on the number of models, which could be (and arguably ought to be) infinite. This is a fundamental problem of, and a recurring theme in, Bayesian model selection.  
\par
As a general rule, problems encountered in parameter estimation tend to be exacerbated in the context of model selection. Computational demands obviously increase with the number of models. Perhaps not so obvious is that many computational schemes that apply straightforwardly in the context of parameter estimation --- such as Markov Chain Monte Carlo sampling \cite{Siekmann:2012ve}, or approximate Bayesian computation (ABC) \cite{Toni:2010p29729,Barnes:2012vs} --- require significant modification and adaptation for model selection.  
\par
As in the case of parameter estimation, appropriate robustness analysis can help. This often involves extending analysis over many related models \cite{Castro:2019ww,Stumpf:2020uh}. In topological sensitivity analysis \cite{Babtie:2014jg}, for example, the structure of the model is perturbed by adding, deleting, or otherwise changing one or more terms in the equations. That is, we go from candidate model $M$ to another model $M'$. We then fit model $M'$, or in reality many alternative models with perturbed structures, to the data to see if alternative models can explain the data. The results of such an analysis can often be sobering, with many alternative model structures being virtually indistinguishable. This level of topological robustness could imply two very different things: (i) the observed behaviour is robust to variation in the model structure; or (ii) the large number of models may simply reflect that the data are not sufficient to discriminate between models. 
 
\subsection{Design Principles}
Even for mathematical modellers in biology the model should be of subsidiary interest compared with the biological phenomena under investigation. Model comparison methods, including model selection, multi-model averaging, and topological sensitivity analysis can be used to group models that provide comparable descriptions of a certain type of biological behaviour, henceforth denoted by $Q$. For biological systems, instances of a behaviour $Q$ could be oscillations, switch-like characteristics \cite{Leon:2016te}, signal filtering \cite{Qiao:2019uw}, robust perfect adaptation \cite{Ma:2009wt,Araujo:2018aa}, or Turing pattern behaviour \cite{Scholes:2019un,Leyshon:2021vc}.
\par
For some types of behaviour, $Q$, it is possible to derive precise mathematical statements that express either the necessary or sufficient conditions (ideally both) for a system to exhibit $Q$. For example, we know that:
\begin{enumerate*}[label=(\roman*)]
  \item multi-stability depends on the presence of positive feedback \cite{Brandman:2005tz};
  \item for robust perfect adaptation (RPA), algebraic conditions have been derived that establish the existence of RPA \cite{Araujo:2018aa};
	\item a Turing instability is required for spontaneous pattern formation in reaction diffusion systems \cite{Scholes:2019un}.
\end{enumerate*}
But generally we still have to validate whether a given model $M$ exhibits behaviour $Q$ or not. Further, for many behaviours $Q$ we may not even have easy criteria to consider.
\par
Given a model universe, $\mathcal{M}$, we can use simulations to identify a subset, $\widetilde{\mathcal{M}} \subseteq \mathcal{M}$, of models that exhibit a particular behaviour $Q$. From the subset $\widetilde{\mathcal{M}}$ we may be able to distill the design principles using a comparative analysis of the models $M \in \widetilde{\mathcal{M}}$. This process, however, is not without difficulties: structural properties, of for example the reaction network, do not always suffice to guarantee the behaviour $Q$; boundary conditions and parameter regimes shape the model output, dynamics, and behaviour, so must also be considered within the comparison methodology. Visual inspection is prohibitive, unreliable, and unrigorous.
\par
The necessity for more flexible, universal, and automatable methodologies to compare models is evident. With this objective in mind, employing simplicial complexes (which are a generalisation of combinatorial graphs) has allowed us to develop flexible representations of both quantitative and qualitative models that are described with any formalism \cite{Vittadello2021b,Vittadello2021c}. Comparison of the simplicial complexes, by our notions of distance or equivalence, provides a meaningful conceptual comparison for the corresponding models. Our model comparison methodology allows for clustering of models according to flexible conceptual considerations, which can lead to the discovery of novel design principles for biological systems.

\subsection{Model Development}
When we cannot rely on first principles for model development we have to use heuristics or intuition to construct models. As we have shown above, this can work but is in no way guaranteed. Considering large model universes, $\mathcal{M}$, and model selection or topological sensitivity analysis offers a potential solution. But every model $M \in \mathcal{M}$ still has to be written down and translated into code to be executed by the computer. When this has to happen for many models, or for very large models, the potential for errors is considerable. If basic properties of models (number and types of interactions, structural properties, kinetics, and so forth) can be specified it is often possible to construct these models automatically, for example by exhaustive enumeration. Metaprogramming provides particularly elegant ways of automatically and on the fly developing and analysing mathematical model structures \cite{Stumpf:2021va}. In metaprogramming, software is considered as part of the data and can therefore be manipulated during run-time and on the fly.
\par
If we can guide the way in which software `writes and rewrites itself', for example by combining it with model selection or robustness analysis, we could take steps towards developing larger and larger, including whole cell, models. In this framework we would start from a model seed, $M^{(0)}$, and propose a new model according to 
\be
M^{(i)} \xlongrightarrow{G(M^{(i)})} M^{(i+1)},
\label{eq:modelseq}
\ee
where $G(M^{(i)})$ is used to propose an update to the model that results in a new model $M^{(i+1)}$; this could, for example, be a stochastic context-free grammar. 
\par
In addition to being, we would argue, less prone to errors than manual curation, this would also facilitate reproducible model development. We already have the ability to share models between groups using file exchange formats \cite{Le-Novere:2005to,Waltemath:2011vk}. But SBML and CellML model specifications do not allow us to deduce how and on what basis certain modelling choices were made; even reading the original publications rarely suffices. If models are created algorithmically then the choices that go into a model-development pipeline can be summarised and communicated efficiently. The development of the model thus becomes reproducible if the updating mechanism and the initial model, $M^{(0)}$, are known.
\par
The procedure in Equation~\eqref{eq:modelseq} can be interpreted as constituting a type of random walk through model space. The process needs to be guided by model selection after every iteration of Equation~\eqref{eq:modelseq}, or every few iterations. The statistical challenges are considerable: (i) conventional parameter estimation at each step would quickly become prohibitive; and (ii) the dimension of the model space can change at each step which makes model selection --- the statistical comparison of successive models, $M^{(i)}$ and $M^{(i+1)}$ --- challenging. Techniques such as reversible jump Markov chain Monte Carlo are required to construct appropriate samplers that can explore varying dimensional spaces.

\subsection{Multiscale, Hierarchical, and Spatio-Temporal Models}
Most biological problems intrinsically span scales: from molecules to cells; from cells to tissues and whole organisms; from species to ecosystems; from genetic variants to populations. Dynamics at one scale have repercussions at other scales, both higher and lower. Sometimes we can only observe data at a higher level and have to draw inferences at the lower level: molecular dynamics drive cell-fate decisions; cellular processes lead to tissue formation; within-host dynamics drive population-level epidemics. The literature in this field is enormous and rapidly growing; we therefore limit the discussion in this section to two aspects: (i) mathematical and statistical challenges related to development and analysis of multiscale models; and (ii) a brief outline of bond graphs as a flexible framework for thermodynamically correct models.
\par
The joint consideration of biological processes across scales opens up new possibilities to capture complex processes. Often we can take a pragmatic approach and use coarse descriptions of the lower-scale dynamics to explore the resulting behaviour at a higher scale. For example, we can use simplified cellular dynamics to model the formation of patterns in tissues. In this manner, agent-based models, for example, allow us to test our hypotheses about intra-cellular dynamics by comparisons with the behaviour observed at the tissue level. The critical evaluation of agreement between multiscale models and biological reality requires careful consideration: just like for models at the same scale there may be many alternative model realisations that give rise to the same behaviour. Testing the sensitivity to model details, a multiscale modelling equivalent of topological sensitivity analysis, does not seem to be widely used. Agent-based models can be computationally prohibitively expensive --- this is already a problem for parameter estimation for agent-based models. Here there is room for the development of emulation methods that replace expensive simulation by statistical models.
\par 
One of the challenges of multi-scale and hierarchical models is to describe the interfaces between levels or modules correctly; here thermodynamics imposes tight constrains \cite{Oster:1973vi} that are generally not incorporated in mathematical biology. Bond graphs have been conceived with that in mind \cite{Gawthrop:2016wd}; they are also an area to which Edmund Crampin made major contributions \cite{Gawthrop:2021ww,Gawthrop:2020ua,Cudmore:2021uc,Pan:2021tq,Shahidi:2021wm,Gawthrop:2017tl,Pan:2018vl}. 
\par
In the bond graph formalism, energy and energy flow are tracked explicitly to ensure that reactions are commensurate with the laws of thermodynamics. 
This has two advantages: (i) resulting models incorporate stoichiometry, conservation of energy, and molecule numbers (mass) from the outset; (ii) thermodynamic constraints lead to models that are automatically modular, like real biological systems. Therefore bond graphs offer a rigorous and well defined framework to develop biophysically and thermodynamically consistent models of molecular processes. Here we will build on this work and extend it to develop whole cell models. 
\par
The critical point in model development is to link up sub-models of, for example, gene regulation, metabolism, and the cell cycle. This modularity turns out to be a natural consequence when thermodynamic conservation laws are applied to the component models. Furthermore, the bond graph formalism provides a straightforward pathway for modelling processes at different levels of complexity, according to available knowledge and data. It can be combined with both multi-physics and multiscale approaches to aid model construction. Here multi-physics refers to  models that can capture different physical processes --- such as heat transfer, dynamics of biochemical reactions over space and time, and transport phenomena inside the cells as well as between the cell and the extracellular environment --- in a consistent manner. Bond graphs allow for this, and therefore provide a natural framework to consider situations where cell mechanics, electrophysiology, and metabolic processes need to be considered jointly.

\subsection{Stochastic Nonlinear Dynamics}
The theories of stochastic processes and dynamical systems have developed largely independently and in parallel. Many investigations in mathematical biology have also considered deterministic dynamics with scant reference to stochasticity. Classical work by Waddington, for example, does not consider noise and random dynamics, with the exception of fluctuating environments. Many important processes in ecology, population biology, epidemiology, cell and developmental biology, and physiology, have been studied with great success using purely deterministic descriptions. Stability analysis and bifurcation theory can give us qualitative, even global qualitative, insights into the complex dynamics underlying ecological dynamics or cell cycle regulation.  
\par
Often the average of a stochastic process can be described adequately using deterministic models. Diffusion and reaction-diffusion systems are good examples for which the underlying stochastic dynamics are sufficiently simple and deterministic partial differential equations can model the system sufficiently: solutions to the diffusion equation have been used, for example, to describe ecological migration, tumour growth, and biological pattern formation.
\par
In population and evolutionary genetics, by contrast, stochasticity --- random sampling of alleles/individuals --- has generally provided the accepted null model. The neutral model of evolution, for example, assumes that most genetic variation has no effect on fitness, the expected number of offspring. Historically, the development of statistics, probability theory, and population genetics are tightly linked. But in ecology and cell biology, for example, stochastic analyses have sometimes been conducted as {\em ad hoc} additions to deterministic analyses.
\par
In many biological problems chance plays an important role, as do nonlinearity and feedback. The analysis of these systems is fraught with challenges. For illustration let us consider stochastic differential equations. Because random processes are not differentiable, we write this in the form,
\be
dX = f(X;\theta) dt + g(X;\eta) dW_t,
\ee
where the first term, $f(X;\theta)$, captures the deterministic dynamics and the second term, $g(X;\eta)$, the stochastic contributions to the dynamics; $dW_t$ denotes the Wiener process increment; and $\theta$ and $\eta$ are the parameters of deterministic and noisy dynamics, respectively. For most biological systems, especially biochemical reaction systems, $f(X;\theta)$ and $g(X;\eta)$ are not independent but related. This relationship can be derived from the microscopic dynamics; frequently, however, $g(X;\eta)$ is chosen for convenience alone, and then typically in the form of white noise, therefore $g(X;\eta) = \sigma^2 = \text{constant}$; this type of noise, which is independent of the state variable $X$ is often referred to as ``additive noise'', in contrast to noise that depends explicitly on the state, which we refer to as ``multiplicative noise''.  
\par
The extent to which qualitative features of the deterministic dynamics affect the behaviour of a model once noise enters the equations requires careful consideration and depends a lot on the type and form of the noise function, $g(X;\eta)$. For example, the fact that the deterministic system is bi-stable in no way guarantees that the corresponding stochastic system has a bi-modal distribution over state space. This lack of conservation of qualitative features of dynamical systems --- including number and nature of stationary states, existence of bifurcations, etc, --- as noise is introduced, or more precisely our inability to predict {\em a priori} if qualitative features are conserved upon the introduction of noise, is becoming a limiting factor in many analyses. In the 1970, for example, Ren{\'{e}}\ Thom and colleagues introduced catastrophe theory\cite{Thom:1989aa}, which describes the types of sudden change (catastrophes) that can be encountered in dynamical systems. This found applications in population and developmental biology, but fell quickly out of favour \cite{Arnold:1992wu}. The past two years have experienced renewed interest in catastrophe theory\cite{Camacho-Aguilar:2021tk,rand2021,Saez:2022wl} and stochastic dynamics, if accounted for correctly, can change the qualitative behaviour \cite{Coomer:2022wb}.
 \par
In short, we do not understand anywhere near enough how to model, analyse, and understand complex dynamical systems that are subject to stochastic effects. Conceptually there are many problems that need to be considered: for example, we would really like to understand how the effects of noise are mitigated or attenuated in order to give rise to highly stable and predictive phenotypes. Computationally, there are also a host of important and worthwhile problems to consider. Simulating stochastic dynamics is challenging and costly. Powerful approximations to exact simulation methods exist, and continue to proliferate \cite{Schnoerr:2017uw}, but ultimately we lack reliable approximations that continue to work when dynamics become nonlinear.

\subsection{Natural Selection}
Evolution provides the organising principle for the life sciences. Dobzhansky's statement ``Nothing in biology makes sense except in the light of evolution'' has become a rallying cry for biologists. The more recent version by Michael Lynch \cite{Lynch:2007vw}, ``Nothing in evolution makes sense except in light of population genetics'', makes this more precise as genetic change and selection on new variants occurs at the population level. Where, when and how selection operates has been at the core of evolutionary analyses and discussions \cite{Williams:1992tt}. Here, as in the previous section, there is a tension between deterministic and stochastic drivers, which need to be considered jointly to understand. The deterministic dynamics reflect the effects of natural selection, and the stochastic dynamics are due to the inherent randomness in the demographic and genealogical dynamics \cite{Ewens:2004th}.
\par
Probabilistic arguments have been dominant in the population genetics literature \cite{Ewens:2004th,Wakeley:2004uq}: random sampling of gametes appears to be remarkably effective at describing many real population genetic diversity patterns. Even the neutral model, which assumes that most genetic variants are not subject to natural selection (or, in its modified form, that most genetic variation is neutral or mildly deleterious), has underpinned the bulk of population genetic analysis. Neutrality makes the problem of studying evolutionary processes and data much easier. Most importantly, neutrality is a powerful null model against which we can test our hypotheses on experimental data.
\par
In population dynamics, deterministic approaches have been used to study, for example, pathogen evolution and early pre-biotic evolutionary processes. One argument that has been put forward is that for sufficiently large population sizes the random evolutionary (demographic sampling) effects become less important. The effects of finite population size have also been largely ignored in game theoretic approaches to evolution. 
\par
The discussion of the previous section is, of course, relevant here as well. The tension between, and combined effects of, deterministic and stochastic dynamics are clearly important in understanding evolutionary dynamics. Simple models are useful, but not always enough \cite{Wakeley:2004uq}; as evolutionary arguments continue to increase in importance in very applied and data-rich areas, such as oncology and host-pathogen evolution, there is a need for both nuance and increased precision. In cancer biology the evolutionary characteristics and the disease aetiology have been shown to be tightly linked \cite{Frank:2007ui,Traulsen:2013vv,MacLean:2014uu,Williams:2016vi,Lakatos:2020tr}. These studies have shown how complicated evolutionary processes are, and there are some salient lessons for evolutionary analysis more generally. Other evidence for the diverse molecular factors affecting evolution come from long term evolution experiments \cite{Blount:2018vi,Card:2019ve,Doebeli:2017tz,Lenski:2017vp,Marshall:2022tx}. 
\par
The upshot of the wealth of data, and the complicated interplay between deterministic and stochastic effects in driving evolution, is that we need more sophisticated maps from genotype to phenotype, especially if the view of omnigenic inheritance persists \cite{Boyle:2017uj,Liu:2019vo,Mathieson:2021vy}. Having better understanding of how genes, their expression, and their interactions affect phenotypes may help in resolving long standing questions in both fundamental biology, but also in understanding the causes and mechanisms of complex diseases. Developing mathematical models for whole cells, and linking these in multi-scale models of organisms, is likely to form an important step in that direction.

\subsection{Hybrid and Data Driven Modelling}
\begin{figure}[h]
\centering\includegraphics[width=1\textwidth]{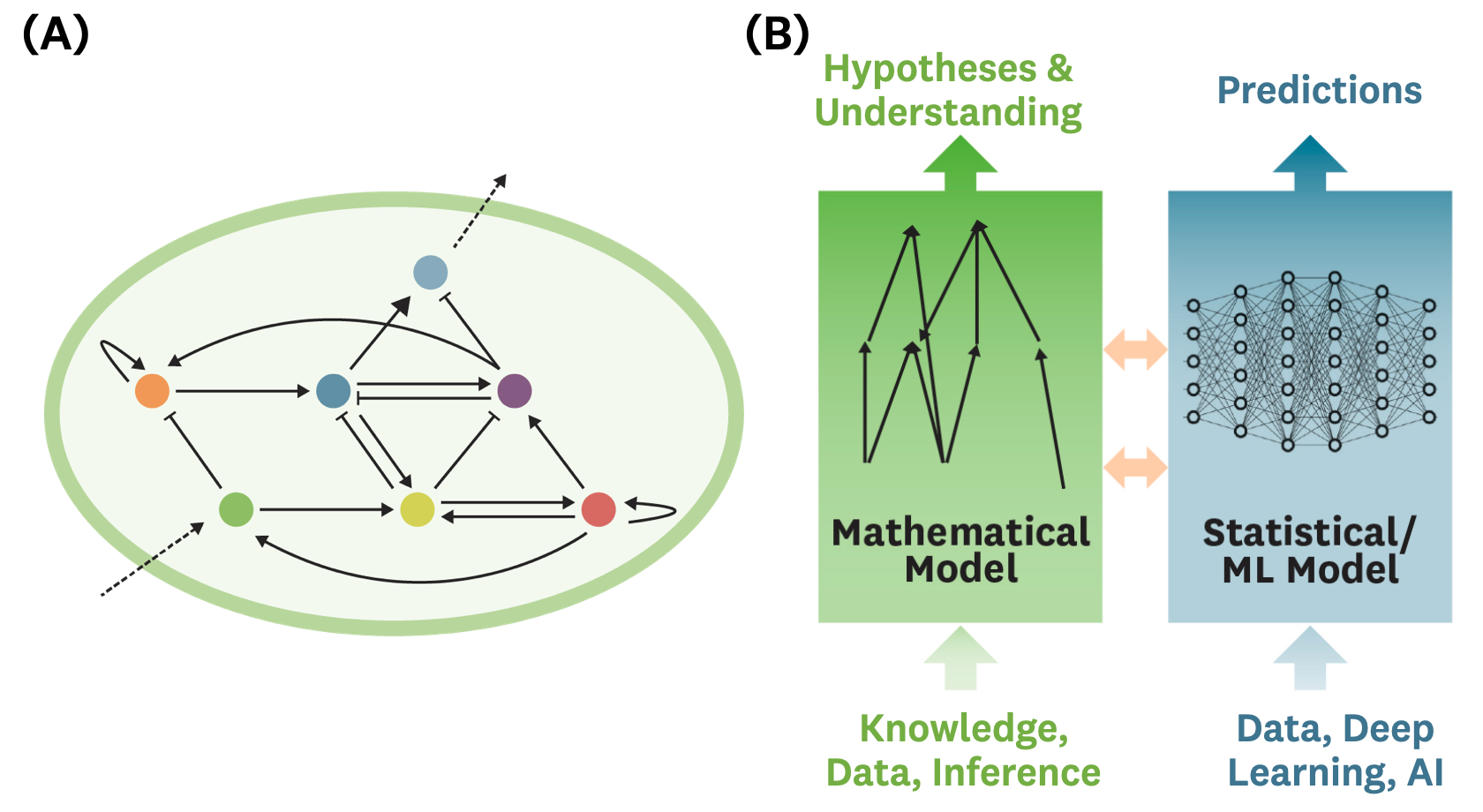}
\caption{(A) We are used to modelling biological systems as, for example, systems of coupled ordinary differential equations. (B) Hybrid models combine statistical models with mechanistic models, and allow investigation of complex systems where only part of the mechanistic relationship is sufficiently well developed to be modelled mechanistcially.}
\label{fig:hybrid}
\end{figure}

Uncertainty about the correct model structure has been a common theme to many of the instances touched upon above. We often lack the knowledge to construct meaningful mathematical models; model selection can help to rule out some models. Automated model development and multi-scale models are opening up new opportunities to increase our universe of mathematical models for biological systems. Hybrid models \cite{Baker:2018vk,Yuan:2021vp}, see Figure~\ref{fig:hybrid}, offer another route towards building better models. 
These models contain mechanistic elements that interface with purely data driven, statistical, or black-box models, to combine the benefits of both approaches. Purely data driven statistical models can provide a baseline against which we compare --- using for example information-theoretic measures such as the Kullback--Leibler divergence \cite{Mc-Mahon:2014aa} --- mechanistic and hybrid models. Further, we can develop data driven (or deductive) modelling approaches as potential emulators to feed back into, for example, the statistical inference approaches where computationally costly simulation is replaced by statistical (``black box'') models \cite{Feng:2022wz}.
\par
Flexibility in choosing between mechanistic and data-driven modelling approaches, and the ability to combine them can open up powerful new ways to study, explain, and predict the behaviour of biological systems. Statistical ``black-box'' and machine learning models, by themselves, provide little direct insight into the underlying causal mechanisms that link genetic components to organismic behaviour. Analysing the resulting models with ``white box'' approaches including saliency maps or input-masked gradients provide important mechanistic, testable, and predictive insights into the critical regulatory regions and their interactions. Knowing the statistical dependencies should ultimately serve as a guide towards generating causal hypotheses for mechanistic/mathematical/functional relationships. We can use hybrid models to test the robustness of learned design principles: the mechanistic model embeds the inferred design principle, and the statistical model the physiological or environmental context (akin to, for example, extrinsic noise in gene expression). 
\par
The concept of neural differential equations \cite{Quaghebeur:2022uj,Roesch:2021ua}, where the right-hand side of the differential equation, $dx/dt = f(x;\theta)$ is replaced by a neural network, 
\be
\frac{dx}{dt} = \text{NN},
\ee
holds the potential to develop new forms of hybrid models, all from within the same dynamical systems perspective.
\par
Both mechanistic modelling and data-driven/machine learning modelling are well developed in their own right. Hybrid modelling, by contrast, is still in its infancy and  the particular combination of hypothesis-driven and data-driven modelling required to develop useful models is only beginning to be studied. There are, however, exciting opportunities to bring methods from, for example, control engineering \cite{Krishnanathan:2012tb,Aquino:2014wt,Lakatos:2016hy} to bear on hybrid modelling attempts. The use of a hybrid-Ansatz in modelling may also resolve some of the outstanding issues in parameter and model identifiability \cite{Villaverde:2016vr,Harrington:2012cr}, where the non-identifiable parts could, or should, perhaps be modelled using a machine learning approach.

\section{A New Mathematical Biology}\label{sec:new}
In this section we change pace and consider some broad mathematical and theoretical ideas for new approaches to mathematical biology. While this discussion is largely from the perspective of pure mathematics, it provides a reflective analysis of the possibilities for the role of mathematics in biology.

\subsection{Overview}
Mathematical biology has predominantly avoided the explicit consideration of life. This paradoxical situation reflects a fundamental deficiency not only of mathematical biology but also, though perhaps to a lesser extent, of theoretical and experimental biology. Of primary concern here is the absence of a general theoretical definition of life and consequent realisation within an appropriate mathematical framework. While progress has been made, including the \emph{relational biology} by theoretical physicist Nicolas Rashevsky \cite{Rashevsky1954} and theoretical biologist Robert Rosen \cite{Rosen1958}, the \emph{hypercycle} by biophysical chemist Manfred Eigen \cite{Eigen1971}, and the \emph{chemoton} by theoretical biologist Tibor G{\'{a}}nti \cite{Ganti2003}, such a key concern is inexplicably the province of only a select few.
\par
Definitions serve to identify and provide the foundation for explanation and knowledge. That a general definition of life is wanting is attendant to the complexity of the emergent phenomenon of life, though arguably more so to the methodological reductionism that has dominated scientific practice, particularly within biology and related fields, in recent decades. Reductionism has been preeminent in experimental biology, and consequently biologists collect enormous quantities of experimental data often without an integrated perspective of the living organism, thus without progression towards knowledge of the phenomenon of life \cite{Cornish_Bowden2011}.
\par
While the ultimate ambition of biology is a general theory of life, this remains an open problem: even general concepts and principles are absent \cite{Cleland2013}. Mathematical biology, along with theoretical biology, can assume a fundamental role in developing a theory of life. Much of current mathematical biology, however, is subject to the same shortcomings as experimental biology. To make progress mathematical biology may need to consider three main aspects, namely emergent behaviour, mathematical maturity, and unification, which together form a basis for a more functional and effective approach to the study of living organisms with mathematics.

\subsection{More is different}
In 1972, Philip W. Anderson\footnote{In 1977 P.W. Anderson shared the Nobel Prize in Physics with Edmund's maternal grand father, Sir Nevill F. Mott, and J.H. Van Vleck.} published his article ``More is different'' \cite{PWAnderson1972}, which became a highly influential exposition on the limitations of methodological reductionism and the importance of emergence in understanding complex phenomena. Anderson relates the hierarchical progression from simplicity to complexity with the progression from reduction to emergence \cite[Chapter II, page 110]{PWAnderson2011}. New levels of system complexity are then associated with the emergence of novel phenomena, concepts, principles, and functions that are consistent with, though not readily explained by, the lower levels of the system. While emergent properties are often associated with the hierarchical, or multiscale, structure of a system \cite[Chapter III, page 139]{PWAnderson2011}, there are alternative perspectives that attempt to avoid any potential circularity arising from this association \cite{Ryan2007}.

Reductionism, where the intention is to explain phenomena via upward causation in relation to underlying processes, has been very effective in science. The limitations of reductionism, however, are readily apparent for both simple and complex systems. For example, copper metal has the properties of being red, shiny, malleable, ductile, and conductive, while individual copper atoms exhibit none of these properties that emerge when many copper atoms form the metal structure \cite[Chapter III, page 143]{PWAnderson2011}. Superconductivity provides an example of emergence in a more complex system: detailed computation of the motions of all electrons and ions in a superconducting material could describe the known phenomenology of superconductivity, however the computation cannot reveal the cause of superconductivity, namely the phenomenon of symmetry breaking. This is simply not captured by the framework at this lower physical level \cite[Chapter III, page 136]{PWAnderson2011}. At much greater complexity, the behaviour of a living cell is an emergent property of the enormous number of molecules of which it consists \cite{IrunRCohen2007}. Cells are regarded as the smallest functional unit of life, a property not associated with, and not readily predictable from, their constituent molecules.

\begin{figure}
\centering\includegraphics[width=1\textwidth]{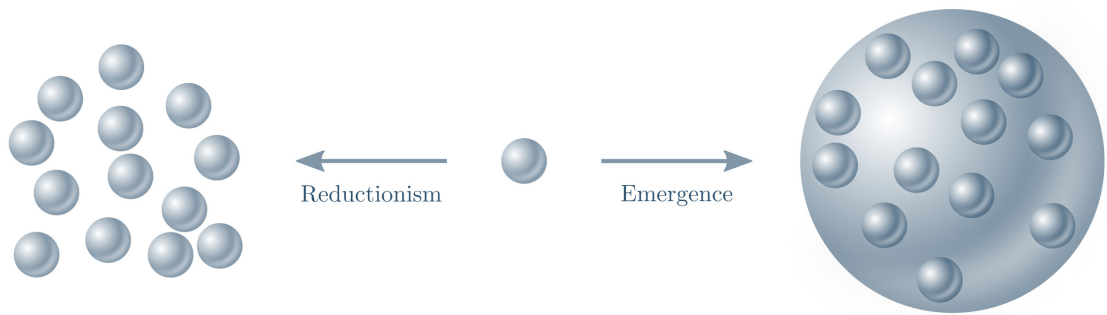}
\caption{Illustration of methodological reductionism and weak emergence. Reductionism assumes that all complex behaviour can be described from knowledge of the underlying constituents. While weak emergence is consistent with the properties of the underlying constituents the emergent behaviour, which may correspond to novel principles of structure, organisation, or function, is not readily described from knowledge of the individual constituents.}
\label{fig:ReductEmerge}
\end{figure}

Different types of emergence have been defined, including pragmatic, nominal, weak, and strong emergence. Strong emergence describes the appearance of an emergent phenomenon that cannot be described with a reductive account, and may be associated with downward causation and emergent forces or physical laws. Weaker forms of emergence are not necessarily associated with novel forces or physical laws. In particular, weak emergence describes a phenomenon that arises in an unexpected or very complicated manner at a higher level of complexity, and would not likely be evident without direct observation, even though a reductive understanding of the physics of the lower levels is in principle sufficient to describe the phenomenon \cite{Bedau2008}. A proposition for the existence of an emergent phenomenon must be accompanied by a rigorous explanation, via a detailed physical mechanism at the level of emergence, for the qualitative difference between the emergent phenomenon and the underlying levels. In Figure~\ref{fig:ReductEmerge} we illustrate the complementary perspectives of methodological reductionism and weak emergence.

A fundamentally important reason for identifying weak emergence is that it may allow for the explanation of a phenomenon without the need to consider the reductive understanding of the lower levels, thereby allowing for a much simplified mathematical description or simulation. Identifying emergence thereby provides for potential flexibility and simplification in describing a complex system. For example, we can consider the motility of a living cell without knowing the movements of all constituent molecules.

While the concept of emergence has been incorporated into many areas of physics, the same cannot be said for biology and related fields, notably mathematical biology. Biologists generally believe that, given a sufficient understanding of chemistry, molecular biology, and cell biology, a complete explanation of living organisms will be achieved through reductive investigation. This is particularly evident in genetics and related areas where the tendency is to believe that understanding gene or protein sequences will lead to a complete explanation of life. To the contrary, the biologist Lynn J. Rothschild illustrates the numerous examples of form and function in biological systems for which the concept of emergence may, if not should, be employed \cite[Chapter 6, page 151]{Clayton2006}. Moreover, the evolutionary biologist Ernst Mayr recognised the ubiquity of emergent behaviour in biological systems, and therefore included emergence as one of six reasons underlying the difficulties in predicting the behaviour of biological systems \cite[Chapter 2, pages 58--59]{Mayr1982}.

While reductionism has been very successful in yielding knowledge of many biochemical and cellular processes underlying biological systems, the limitations of reductionism are now apparent from the seemingly underestimated complexity of these systems \cite{Van_Regenmortel2004}. By identifying emergent phenomena we can explain the properties of a biological system without the limitations of accounting for the complexity of the underlying constituents. Whether or not living organisms can be explained in principle from biochemistry, and thereby the laws of physics, is not of concern from the pragmatic perspective of utilising emergence to simplify and expedite the process of understanding biological systems.

Embracing emergence in biology does not dispel the need for much of current mathematical biology, but requires the additional consideration of identifying emergent phenomena at higher levels of system complexity. The latter is a very difficult program that requires new theoretical and mathematical ideas, some of which we have outlined above. The result, however, will be a much more sophisticated understanding and description of biological systems that yields mathematical models which avert prohibitively complicated computations. For this, rigorous mathematical definitions of emergence and emergent properties are required, and while there are some initial considerations \cite{Bar_Yam2004,Ryan2007}, further work is required.

To summarise with the words of Anderson on the ``triumph of emergence over reductionism'': ``... large objects such as ourselves are, in myriad ways, the product of principles of organization and of collective behavior which in no meaningful sense can be reduced to the behavior of our elementary constituents. Large objects are often more constrained by those principles than by what the principles act upon.'' \cite[Chapter 4, Pages 195--196]{PWAnderson2011}.

\subsection{Mathematical Maturity}
The list of articles in any issue of the Journal of Mathematical Physics illustrates the mathematical maturity enjoyed by theoretical physics, notably the tendency for mathematical rigour including the use of the definition-theorem-proof style of pure mathematics \cite{Thurston1994}, and the use of sophisticated mathematical theory. Currently, only a relatively small proportion of the research literature in mathematical biology exhibits such maturity.

Nicolas Rashevsky, one of the pioneers of mathematical biology, envisioned that the relationship between mathematical biology and experimental biology would correspond to that between mathematical physics and experimental physics \cite{Cull2007}. That the vision of Rashevsky has been realised in such limited ways is probably due to several reasons, which we have already touched on above: biological systems are generally more complex than inanimate systems; avoidance by mathematicians who regard biology as unsophisticated and lacking rigorous definitions; strictly reductionist approaches, possibly influenced by experimental biology; insufficiency of deep theorising about living organisms, and in particular the formation of rigorous definitions; imprudent importation of mathematical models from other fields, such as physics and engineering; complacency with modelling simple biological systems with simple models and simple mathematics; indifference, or inability, toward applying more sophisticated mathematics or developing new mathematics.

The complexity of living organisms relative to physical systems suggests the need for the application of mathematical theory that is even more sophisticated, as well as the need for new mathematical ideas leading to new areas of mathematics \cite{cohen2004,Reed2015,Borovik2021}. For this, rigorous definitions of biological concepts are a necessity, and without such rigour the progression of mathematical biology, and how it supports experimental biology, is clearly limited. Mathematical biology requires new mathematical theory \cite{Borovik2021}, and the passive importation of mathematical models from other fields, particularly physics and engineering, and their application in superficial ways to biological systems runs counter to the constructive progression of mathematical and theoretical biology. Capturing the intricacies of living systems, especially their emergent properties, will require but also inspire such new mathematics.
%

There is, we feel, increasing need of employing the definition-theorem-proof style of mathematics in mathematical biology. 
The general lack of rigorous definitions in biology is of fundamental concern for mathematical biology since all mathematical assertions originate from definitions, and the requirement to state clear definitions in mathematical biology would stimulate much needed theoretical thought regarding biological concepts. Further, a theorem asserts a general principle, and formulating a mathematical result as a theorem encourages the discovery of more general ideas rather than individual cases, demonstrates rigour, allows for immediate connection with other mathematical theory, simplifies the exposition, and may occasion a deeper understanding of the mathematics and the corresponding system. Rigorous mathematics, consisting of the definition-theorem-proof style is indubitably of importance for the success of mathematical biology \cite{Clairambault2013}.

\subsection{Unification}
Given the lack of general theoretical frameworks in biology \cite{Scheiner2010} it is of no surprise that biological knowledge is fragmented, even to the extent that different subfields routinely use distinct terms for the same concept \cite{Blagosklonny2002,Barnes2002}. In principle, mathematical and theoretical biology should be placed to remedy this situation, however they are arguably in a similar position which may only serve to further this fragmentation. The fundamental need for unification, or integration, within mathematical biology has long been realised, however little progress has been made. Here we discuss three particular areas associated with mathematical biology for which unification is imperative: biological data, mathematical models, and system perspective.

In his 2002 Nobel lecture, molecular biologist Sydney Brenner said that ``We are all conscious today that we are drowning in a sea of data and starving for knowledge'' \cite{Brenner2002}. Rather than mere descriptions and routine analysis of biological data, deep theorising and integration of the often heterogeneous data is required to yield knowledge \cite{Nurse2021}. Mathematical modelling can be of much assistance here, particularly regarding the assessment of consistency between a hypothesis and the experimental data, or for the identification of design principles. Mathematical models are fundamental for developing knowledge of an observed biological phenomenon, such as the identification of design principles. Given a causative hypothesis for a phenomenon we can construct a mathematical model that represents the hypothesis explicitly, and then compare model predictions with experimental data to determine whether the hypothesis is consistent, or not, with the evidence. In opposition to much of current practice in mathematical biology, however, integration of heterogeneous data requires the consideration of multiple insightful hypotheses, and therefore multiple models, or the consideration of a single hypothesis for consistency with a broad range of experimental data. Consideration of a single hypothesis/model within the context of a single data set, or multiple data sets that are closely related, achieves little toward data integration and novel biological theory. The objective is the development of new theoretical knowledge: being able to simulate a phenomenon does not in itself correspond to an explanation of the phenomenon.

The unification of mathematical models is at the frontier of mathematical biology, and general methodologies for comparing the underlying conceptual structure of models, irrespective of mathematical formalism, provide essential facilitation for the process \cite{Vittadello2021b,Vittadello2021c}. Model comparison can partition a set of models into categories where similar models, regarded as sharing an essential qualitative characteristic, belong to the same category. Similar models may simply be incremental variations of each other, or may originally appear distinct and unrelated. Discovering similarities between seemingly distinct models can lead to novel theoretical insights into biological phenomena \cite{Vittadello2021b}, and ultimately to a unified mathematical description of a biological system. A perspective of unification can encourage the avoidance of excessive incrementalism in model development and, along with a rigorous definition-theorem-proof style, yield general mathematical frameworks for biological systems, rather than the essentially algorithmic approach of developing simple models of simple systems.

It is widely recognised that to understand a biological phenomenon it must be perceived as an integrated whole \cite{Rashevsky1954}. For example, a biological function has meaning for an organism at the level at which the function is integrated, and below this level the function does not exist \cite{Noble2011,Noble2012,Rajapakse2017}. Ecology has, of course, long recognised this. Comprehending the integration is particularly difficult since at a given level of organisation in a biological system the biological constituents are generally heterogeneous \cite{cohen2004}, and naturally span several scales. Most of current mathematical biology is either concerned with ``life-less'' systems such as biochemical networks, or considers living organisms (such as a cell) as inanimate objects. 
The latter is a pragmatic and justified choice when the aim is to develop models that replicate, rather than explain, the behaviour of living systems, for example cell migration in response to gradients. But it falls way short of describing essential hallmarks of life. Here there appears to be genuine scope for a new mathematical biology that captures such hallmarks, including life as an emergent phenomenon.


\section{Conclusion}
Open problems in research can be interpreted as gaping holes in our understanding, or as exciting opportunities for future research. Most of the problems discussed here have aspects of both. There are considerable technical and computational challenges, but also some conceptual challenges of a mathematical nature, related to model development and model calibration. There is a definite appeal of using and combining methods from dynamical systems theory, computer science, statistical inference and machine learning to develop better, more reproducible methods of model development. Multiscale modelling, hybrid models, model selection, and automated generation of models individually, but especially when taken together, offer enormous potential to expand the scope of mathematical biology. We believe that the ability to explain, predict and control living systems requires a broadening of the toolbox of mathematical biologists.
\par
In our view the appeal of combining hypothesis-driven and data-driven modelling frameworks is considerable. The technical challenges in doing so are also considerable. Data-driven modelling has seen a rapid rise over the past decade and deep neural networks have become powerful tools in biology \cite{Sapoval:2022wo,AlQuraishi:2021wr,Naert:2021wc}; and their predictive performance is perhaps more immediately persuasive than are the mechanistic insights that modelling can provide. But both are needed \cite{Nurse2021}.
\par
We believe that a new mathematical biology will be required to understand life at the fundamental level we discussed in the previous section. Since this requires the substantial development of new biological and mathematical theory, we have little concrete idea of what this will look like. Anderson's work \cite{PWAnderson1972}, maybe coupled to thermodynamic considerations \cite{Oster:1973vi}, provides a starting perspective: emergent phenomena, including life itself, are not readily explained from principles that are detected by reductionist approaches.
\section*{Acknowledgements}
The authors thank the members of the Theoretical Systems Biology group, past and present, for stimulating discussions that have shaped our view. Much of the conceptual work has been the result of our ongoing collaborations, especially in MACSYS. Both authors acknowledge the financial support from the Volkswagen Foundation through a {\em Life?} program grant. 
\clearpage
\bibliographystyle{unsrt}
\bibliography{mst_refs,SeanBibliography}

\newcommand{\noop}[1]{}
\begin{thebibliography}{100}

\bibitem{May:2004p11952}
May.
\newblock {Uses and abuses of mathematics in biology.}
\newblock {\em Science}, 303:790, 2004.

\bibitem{cohen2004}
Cohen.
\newblock Mathematics is biology's next microscope, only better; biology is
  mathematics' next physics, only better.
\newblock {\em PLoS Biol}, 2(12):e439, 2004.

\bibitem{Bizzarri:2019aa}
Mariano Bizzarri, Douglas~E Brash, James Briscoe, Ver{\^o}nica~A Grieneisen,
  Claudio~D Stern, and Michael Levin.
\newblock A call for a better understanding of causation in cell biology.
\newblock {\em Nat Rev Mol Cell Biol}, 20(5):261--262, 2019.

\bibitem{Cursons:2015wf}
Joseph Cursons, Jerry Gao, Daniel~G Hurley, Cristin~G Print, P~Rod Dunbar,
  Marc~D Jacobs, and Edmund~J Crampin.
\newblock Regulation of erk-mapk signaling in human epidermis.
\newblock {\em BMC Syst Biol}, 9:41, Jul 2015.

\bibitem{Filippi:2016gs}
S~Filippi, Chris~P Barnes, P~D~W Kirk, Takamasa Kudo, Katsuyuki Kunida,
  Siobhan~S McMahon, Takaho Tsuchiya, Takumi Wada, Shinya Kuroda, and Michael
  P~H Stumpf.
\newblock {Robustness of MEK-ERK Dynamics and Origins of Cell-to-Cell
  Variability in MAPK Signaling}.
\newblock {\em Cell reports}, 15(11):2524--2535, 2016.

\bibitem{Thorne:2017tm}
Kip~S. Thorne and Roger~D. Blandford.
\newblock {\em Modern classical physics: optics, fluids, plasmas, elasticity,
  relativity, and statistical physics}.
\newblock Princeton University Press, 2017.

\bibitem{Weyl:1952uz}
Hermann Weyl.
\newblock {\em Symmetry}.
\newblock Princeton University Press, Princeton, 1952.

\bibitem{Goldstein:2002up}
Herbert Goldstein, Charles~P Poole, and John~L Safko.
\newblock {\em Classical mechanics}.
\newblock Addison Wesley, San Francisco, 3rd ed edition, 2002.

\bibitem{PWAnderson1972}
P.~W. Anderson.
\newblock More is different.
\newblock {\em Science}, 177:393--396, 1972.

\bibitem{MacArthur:2022ww}
Ben~D MacArthur.
\newblock The geometry of cell fate.
\newblock {\em Cell Syst}, 13(1):1--3, 01 2022.

\bibitem{Kirk:2015gj}
Paul D~W Kirk, Ann~C Babtie, and Michael P~H Stumpf.
\newblock {Systems biology (un)certainties.}
\newblock {\em Science}, 350(6259):386, 2015.

\bibitem{Villaverde:2019ws}
Alejandro~F Villaverde, Fabian Fr{\"o}hlich, Daniel Weindl, Jan Hasenauer, and
  Julio~R Banga.
\newblock Benchmarking optimization methods for parameter estimation in large
  kinetic models.
\newblock {\em Bioinformatics}, 35(5):830--838, 03 2019.

\bibitem{Villaverde:2016vr}
Alejandro~F Villaverde, Antonio Barreiro, and Antonis Papachristodoulou.
\newblock Structural identifiability of dynamic systems biology models.
\newblock {\em PLoS Comput Biol}, 12(10):e1005153, Oct 2016.

\bibitem{Gabor:2015wd}
Attila G{\'a}bor and Julio~R Banga.
\newblock Robust and efficient parameter estimation in dynamic models of
  biological systems.
\newblock {\em BMC Syst Biol}, 9:74, Oct 2015.

\bibitem{Liepe:2014iwa}
Juliane Liepe, P~D~W Kirk, S~Filippi, T~Toni, Chris~P Barnes, and Michael P~H
  Stumpf.
\newblock {A framework for parameter estimation and model selection from
  experimental data in systems biology using approximate Bayesian computation}.
\newblock {\em Nature Prot}, 9(2):439--456, 2014.

\bibitem{Toni:2009tr}
Toni, Welch, Strelkowa, Ipsen, and Stumpf.
\newblock {Approximate Bayesian computation scheme for parameter inference and
  model selection in dynamical systems.}
\newblock {\em J Roy Soc Interface}, 6:187, 2009.

\bibitem{Komorowski:2011cn}
M~Komorowski, Maria~J Costa, David~A Rand, and Michael P~H Stumpf.
\newblock {Sensitivity, robustness, and identifiability in stochastic chemical
  kinetics models.}
\newblock {\em Proc Natl Acad Sci}, 108(21):8645--8650, 2011.

\bibitem{Secrier:2009tu}
Maria Secrier, Tina Toni, and Michael P~H Stumpf.
\newblock The abc of reverse engineering biological signalling systems.
\newblock {\em Mol Biosyst}, 5(12):1925--35, Dec 2009.

\bibitem{Daniels:2008vr}
Bryan~C Daniels, Yan-Jiun Chen, James~P Sethna, Ryan~N Gutenkunst, and
  Christopher~R Myers.
\newblock Sloppiness, robustness, and evolvability in systems biology.
\newblock {\em Curr Opin Biotechnol}, 19(4):389--95, Aug 2008.

\bibitem{Liepe:2013vj}
Juliane Liepe, Sarah Filippi, Micha{\l} Komorowski, and Michael P~H Stumpf.
\newblock Maximizing the information content of experiments in systems biology.
\newblock {\em PLoS Comput Biol}, 9(1):e1002888, 2013.

\bibitem{Babtie:2017ix}
Ann~C Babtie and Michael~P.H. Stumpf.
\newblock {How to deal with parameters for whole-cell modelling.}
\newblock {\em J Roy Soc Interface}, 14(133):20170237, 2017.

\bibitem{Nam:2020td}
Kee-Myoung Nam, Benjamin~M Gyori, Silviana~V Amethyst, Daniel~J Bates, and
  Jeremy Gunawardena.
\newblock Robustness and parameter geography in post-translational modification
  systems.
\newblock {\em PLoS Comput Biol}, 16(5):e1007573, 05 2020.

\bibitem{Kirk:2013hq}
Paul D~W Kirk, Thomas~W Thorne, and Michael P~H Stumpf.
\newblock {Model selection in systems and synthetic biology.}
\newblock {\em Current opinion in biotechnology}, 24(4):767, 2013.

\bibitem{Barnes:2011hh}
Chris~P Barnes, Daniel Silk, X~Sheng, Michael P~H Stumpf, and Michael P~H
  Stumpf.
\newblock {Bayesian design of synthetic biological systems.}
\newblock {\em Proc Natl Acad Sci USA}, 108(37):15190--15195, 2011.

\bibitem{Siekmann:2012ve}
Ivo Siekmann, James Sneyd, and Edmund~J Crampin.
\newblock Mcmc can detect nonidentifiable models.
\newblock {\em Biophys J}, 103(11):2275--86, Dec 2012.

\bibitem{Toni:2010p29729}
Tina Toni and Michael P~H Stumpf.
\newblock {Simulation-based model selection for dynamical systems in systems
  and population biology.}
\newblock {\em Bioinformatics}, 26:104, 2010.

\bibitem{Barnes:2012vs}
Chris~P Barnes, Sarah Filippi, Michael P~H Stumpf, and Thomas~W Thorne.
\newblock Considerate approaches to constructing summary statistics for abc
  model selection.
\newblock {\em Stat and Comp}, 22:1181--1197, 2012.

\bibitem{Castro:2019ww}
Dayanne~M Castro, Nicholas~R de~Veaux, Emily~R Miraldi, and Richard Bonneau.
\newblock Multi-study inference of regulatory networks for more accurate models
  of gene regulation.
\newblock {\em PLoS Comput Biol}, 15(1):e1006591, 01 2019.

\bibitem{Stumpf:2020uh}
Michael P~H Stumpf.
\newblock Multi-model and network inference based on ensemble estimates:
  avoiding the madness of crowds.
\newblock {\em J R Soc Interface}, 17(171):20200419, 10 2020.

\bibitem{Babtie:2014jg}
Ann~C Babtie, Paul D~W Kirk, and Michael P~H Stumpf.
\newblock {Topological sensitivity analysis for systems biology.}
\newblock {\em Proc Natl Acad Sci}, 111(52):18507, 2014.

\bibitem{Leon:2016te}
Miriam Leon, Mae~L Woods, Alex J~H Fedorec, and Chris~P Barnes.
\newblock A computational method for the investigation of multistable systems
  and its application to genetic switches.
\newblock {\em BMC Syst Biol}, 10(1):130, 12 2016.

\bibitem{Qiao:2019uw}
Lingxia Qiao, Wei Zhao, Chao Tang, Qing Nie, and Lei Zhang.
\newblock Network topologies that can achieve dual function of adaptation and
  noise attenuation.
\newblock {\em Cell Syst}, 9(3):271--285.e7, 09 2019.

\bibitem{Ma:2009wt}
Wenzhe Ma, Ala Trusina, Hana El-Samad, Wendell~A Lim, and Chao Tang.
\newblock Defining network topologies that can achieve biochemical adaptation.
\newblock {\em Cell}, 138(4):760--73, Aug 2009.

\bibitem{Araujo:2018aa}
Robyn~P Araujo and Lance~A Liotta.
\newblock The topological requirements for robust perfect adaptation in
  networks of any size.
\newblock {\em Nat Commun}, 9(1):1757, 2018.

\bibitem{Scholes:2019un}
Natalie~S Scholes, David Schnoerr, Mark Isalan, and Michael P~H Stumpf.
\newblock A comprehensive network atlas reveals that turing patterns are common
  but not robust.
\newblock {\em Cell Syst}, 9(3):243--257.e4, 09 2019.

\bibitem{Leyshon:2021vc}
Thomas Leyshon, Elisa Tonello, David Schnoerr, Heike Siebert, and Michael P~H
  Stumpf.
\newblock The design principles of discrete turing patterning systems.
\newblock {\em J Theor Biol}, 531:110901, 12 2021.

\bibitem{Brandman:2005tz}
Onn Brandman, James~E Ferrell, Jr, Rong Li, and Tobias Meyer.
\newblock Interlinked fast and slow positive feedback loops drive reliable cell
  decisions.
\newblock {\em Science}, 310(5747):496--8, Oct 2005.

\bibitem{Vittadello2021b}
Sean~T. Vittadello and Michael P.~H. Stumpf.
\newblock Model comparison via simplicial complexes and persistent homology.
\newblock {\em Royal Society Open Science}, 8:211361, 2021.

\bibitem{Vittadello2021c}
Sean~T. Vittadello and Michael P.~H. Stumpf.
\newblock A group theoretic approach to model comparison with simplicial
  representations.
\newblock {\em arXiv:2111.02170v1}, 2021.

\bibitem{Stumpf:2021va}
Michael P~H Stumpf.
\newblock Statistical and computational challenges for whole cell modelling.
\newblock {\em Current Opinion in Systems Biology}, 26:58--63, 2021.

\bibitem{Le-Novere:2005to}
Nicolas Le~Nov{\`e}re, Andrew Finney, Michael Hucka, Upinder~S Bhalla, Fabien
  Campagne, Julio Collado-Vides, Edmund~J Crampin, Matt Halstead, Edda Klipp,
  Pedro Mendes, Poul Nielsen, Herbert Sauro, Bruce Shapiro, Jacky~L Snoep,
  Hugh~D Spence, and Barry~L Wanner.
\newblock Minimum information requested in the annotation of biochemical models
  (miriam).
\newblock {\em Nat Biotechnol}, 23(12):1509--15, Dec 2005.

\bibitem{Waltemath:2011vk}
Dagmar Waltemath, Richard Adams, Daniel~A Beard, Frank~T Bergmann, Upinder~S
  Bhalla, Randall Britten, Vijayalakshmi Chelliah, Michael~T Cooling, Jonathan
  Cooper, Edmund~J Crampin, Alan Garny, Stefan Hoops, Michael Hucka, Peter
  Hunter, Edda Klipp, Camille Laibe, Andrew~K Miller, Ion Moraru, David
  Nickerson, Poul Nielsen, Macha Nikolski, Sven Sahle, Herbert~M Sauro, Henning
  Schmidt, Jacky~L Snoep, Dominic Tolle, Olaf Wolkenhauer, and Nicolas
  Le~Nov{\`e}re.
\newblock Minimum information about a simulation experiment (miase).
\newblock {\em PLoS Comput Biol}, 7(4):e1001122, Apr 2011.

\bibitem{Oster:1973vi}
G~F Oster, A~S Perelson, and A~Katchalsky.
\newblock Network thermodynamics: dynamic modelling of biophysical systems.
\newblock {\em Q Rev Biophys}, 6(1):1--134, Feb 1973.

\bibitem{Gawthrop:2016wd}
Peter~J Gawthrop and Edmund~J Crampin.
\newblock Modular bond-graph modelling and analysis of biomolecular systems.
\newblock {\em IET Syst Biol}, 10(5):187--201, Oct 2016.

\bibitem{Gawthrop:2021ww}
Peter~J Gawthrop, Michael Pan, and Edmund~J Crampin.
\newblock Modular dynamic biomolecular modelling with bond graphs: the
  unification of stoichiometry, thermodynamics, kinetics and data.
\newblock {\em J R Soc Interface}, 18(181):20210478, 08 2021.

\bibitem{Gawthrop:2020ua}
Peter~J Gawthrop, Peter Cudmore, and Edmund~J Crampin.
\newblock Physically-plausible modelling of biomolecular systems: A simplified,
  energy-based model of the mitochondrial electron transport chain.
\newblock {\em J Theor Biol}, 493:110223, 05 2020.

\bibitem{Cudmore:2021uc}
Peter Cudmore, Michael Pan, Peter~J Gawthrop, and Edmund~J Crampin.
\newblock Analysing and simulating energy-based models in biology using
  bondgraphtools.
\newblock {\em Eur Phys J E Soft Matter}, 44(12):148, Dec 2021.

\bibitem{Pan:2021tq}
Michael Pan, Peter~J Gawthrop, Joseph Cursons, and Edmund~J Crampin.
\newblock Modular assembly of dynamic models in systems biology.
\newblock {\em PLoS Comput Biol}, 17(10):e1009513, 10 2021.

\bibitem{Shahidi:2021wm}
Niloofar Shahidi, Michael Pan, Soroush Safaei, Kenneth Tran, Edmund~J Crampin,
  and David~P Nickerson.
\newblock Hierarchical semantic composition of biosimulation models using bond
  graphs.
\newblock {\em PLoS Comput Biol}, 17(5):e1008859, 05 2021.

\bibitem{Gawthrop:2017tl}
Peter~J Gawthrop and Edmund~J Crampin.
\newblock Energy-based analysis of biomolecular pathways.
\newblock {\em Proc Math Phys Eng Sci}, 473(2202):20160825, Jun 2017.

\bibitem{Pan:2018vl}
Michael Pan, Peter~J Gawthrop, Kenneth Tran, Joseph Cursons, and Edmund~J
  Crampin.
\newblock Bond graph modelling of the cardiac action potential: implications
  for drift and non-unique steady states.
\newblock {\em Proc Math Phys Eng Sci}, 474(2214):20180106, Jun 2018.

\bibitem{Thom:1989aa}
Rene Thom.
\newblock {\em Structural stability and morphogenesis}.
\newblock Addison-Wesley, 1989.

\bibitem{Arnold:1992wu}
V.~I Arnold.
\newblock {\em Catastrophe theory}.
\newblock Springer-Verlag, Berlin, 3rd rev. and expanded ed edition, 1992.

\bibitem{Camacho-Aguilar:2021tk}
Elena Camacho-Aguilar, Aryeh Warmflash, and David~A Rand.
\newblock Quantifying cell transitions in c. elegans with data-fitted landscape
  models.
\newblock {\em PLoS Comput Biol}, 17(6):e1009034, 06 2021.

\bibitem{rand2021}
Rand, Raju, S{\'a}ez, Corson, and Siggia.
\newblock Geometry of gene regulatory dynamics.
\newblock {\em Proc Natl Acad Sci}, 118(38):e2109729118, 2021.

\bibitem{Saez:2022wl}
Meritxell S{\'a}ez, Robert Blassberg, Elena Camacho-Aguilar, Eric~D Siggia,
  David~A Rand, and James Briscoe.
\newblock Statistically derived geometrical landscapes capture principles of
  decision-making dynamics during cell fate transitions.
\newblock {\em Cell Syst}, 13(1):12--28.e3, 01 2022.

\bibitem{Coomer:2022wb}
Megan~A Coomer, Lucy Ham, and Michael P~H Stumpf.
\newblock Noise distorts the epigenetic landscape and shapes cell-fate
  decisions.
\newblock {\em Cell Syst}, 13(1):83--102.e6, 01 2022.

\bibitem{Schnoerr:2017uw}
David Schnoerr, G~Sanguinetti, and Ramon Grima.
\newblock Approximation and inference methods for stochastic biochemical
  kinetics---a tutorial review.
\newblock {\em Journal of Physics A}, 50(9):093001, 2017.

\bibitem{Lynch:2007vw}
Michael Lynch.
\newblock {\em The origins of genome architecture}.
\newblock Sinauer Associates, Sunderland, Mass., 2007.

\bibitem{Williams:1992tt}
George~C Williams.
\newblock {\em Natural selection: domains, levels, and challenges}.
\newblock Oxford University Press, New York, 1992.

\bibitem{Ewens:2004th}
W.~J Ewens.
\newblock {\em Mathematical population genetics}, volume v. 27.
\newblock Springer, New York, 2nd ed edition, 2004.

\bibitem{Wakeley:2004uq}
J~Wakeley.
\newblock Recent trends in population genetics: more data! more math! simple
  models?
\newblock {\em J Hered}, 95(5):397--405, 2004.

\bibitem{Frank:2007ui}
Steven~A. Frank.
\newblock {\em Dynamics of cancer: incidence, inheritance, and evolution}.
\newblock Princeton University Press, Princeton, N.J., 2007.

\bibitem{Traulsen:2013vv}
Arne Traulsen, Tom Lenaerts, Jorge~M Pacheco, and David Dingli.
\newblock On the dynamics of neutral mutations in a mathematical model for a
  homogeneous stem cell population.
\newblock {\em J R Soc Interface}, 10(79):20120810, Feb 2013.

\bibitem{MacLean:2014uu}
Adam~L MacLean, Sarah Filippi, and Michael P~H Stumpf.
\newblock The ecology in the hematopoietic stem cell niche determines the
  clinical outcome in chronic myeloid leukemia.
\newblock {\em Proc Natl Acad Sci U S A}, 111(10):3883--8, Mar 2014.

\bibitem{Williams:2016vi}
Marc~J Williams, Benjamin Werner, Chris~P Barnes, Trevor~A Graham, and Andrea
  Sottoriva.
\newblock Identification of neutral tumor evolution across cancer types.
\newblock {\em Nat Genet}, 48(3):238--244, Mar 2016.

\bibitem{Lakatos:2020tr}
Eszter Lakatos, Marc~J Williams, Ryan~O Schenck, William C~H Cross, Jacob
  Househam, Luis Zapata, Benjamin Werner, Chandler Gatenbee, Mark
  Robertson-Tessi, Chris~P Barnes, Alexander R~A Anderson, Andrea Sottoriva,
  and Trevor~A Graham.
\newblock Evolutionary dynamics of neoantigens in growing tumors.
\newblock {\em Nat Genet}, 52(10):1057--1066, 10 2020.

\bibitem{Blount:2018vi}
Zachary~D Blount, Richard~E Lenski, and Jonathan~B Losos.
\newblock Contingency and determinism in evolution: Replaying life's tape.
\newblock {\em Science}, 362(6415), 11 2018.

\bibitem{Card:2019ve}
Kyle~J Card, Thomas LaBar, Jasper~B Gomez, and Richard~E Lenski.
\newblock Historical contingency in the evolution of antibiotic resistance
  after decades of relaxed selection.
\newblock {\em PLoS Biol}, 17(10):e3000397, 10 2019.

\bibitem{Doebeli:2017tz}
Michael Doebeli, Yaroslav Ispolatov, and Burt Simon.
\newblock Towards a mechanistic foundation of evolutionary theory.
\newblock {\em Elife}, 6, 02 2017.

\bibitem{Lenski:2017vp}
Richard~E Lenski.
\newblock What is adaptation by natural selection? perspectives of an
  experimental microbiologist.
\newblock {\em PLoS Genet}, 13(4):e1006668, 04 2017.

\bibitem{Marshall:2022tx}
Dustin~J Marshall, Martino Malerba, Thomas Lines, Aysha~L Sezmis, Chowdhury~M
  Hasan, Richard~E Lenski, and Michael~J McDonald.
\newblock Long-term experimental evolution decouples size and production costs
  in escherichia coli.
\newblock {\em Proc Natl Acad Sci U S A}, 119(21):e2200713119, 05 2022.

\bibitem{Boyle:2017uj}
Evan~A Boyle, Yang~I Li, and Jonathan~K Pritchard.
\newblock An expanded view of complex traits: From polygenic to omnigenic.
\newblock {\em Cell}, 169(7):1177--1186, Jun 2017.

\bibitem{Liu:2019vo}
Xuanyao Liu, Yang~I Li, and Jonathan~K Pritchard.
\newblock Trans effects on gene expression can drive omnigenic inheritance.
\newblock {\em Cell}, 177(4):1022--1034.e6, 05 2019.

\bibitem{Mathieson:2021vy}
Iain Mathieson.
\newblock The omnigenic model and polygenic prediction of complex traits.
\newblock {\em Am J Hum Genet}, 108(9):1558--1563, 09 2021.

\bibitem{Baker:2018vk}
Ruth~E Baker, Jose-Maria Pe{\~n}a, Jayaratnam Jayamohan, and Antoine
  J{\'e}rusalem.
\newblock Mechanistic models versus machine learning, a fight worth fighting
  for the biological community?
\newblock {\em Biol Lett}, 14(5), 05 2018.

\bibitem{Yuan:2021vp}
Bo~Yuan, Ciyue Shen, Augustin Luna, Anil Korkut, Debora~S Marks, John Ingraham,
  and Chris Sander.
\newblock Cellbox: Interpretable machine learning for perturbation biology with
  application to the design of cancer combination therapy.
\newblock {\em Cell Syst}, 12(2):128--140.e4, 02 2021.

\bibitem{Mc-Mahon:2014aa}
Siobhan~S Mc~Mahon, Aaron Sim, Sarah Filippi, Robert Johnson, Juliane Liepe,
  Dominic Smith, and Michael P~H Stumpf.
\newblock Information theory and signal transduction systems: from molecular
  information processing to network inference.
\newblock {\em Semin Cell Dev Biol}, 35:98--108, Nov 2014.

\bibitem{Feng:2022wz}
Yongxiang Feng, Zhen Cheng, Huichao Chai, Weihua He, Liang Huang, and Wenhui
  Wang.
\newblock Neural network-enhanced real-time impedance flow cytometry for
  single-cell intrinsic characterization.
\newblock {\em Lab Chip}, 22(2):240--249, 01 2022.

\bibitem{Quaghebeur:2022uj}
Ward Quaghebeur, Elena Torfs, Bernard De~Baets, and Ingmar Nopens.
\newblock Hybrid differential equations: Integrating mechanistic and
  data-driven techniques for modelling of water systems.
\newblock {\em Water Res}, 213:118166, Feb 2022.

\bibitem{Roesch:2021ua}
Elisabeth Roesch, Christopher Rackauckas, and Michael P~H Stumpf.
\newblock Collocation based training of neural ordinary differential equations.
\newblock {\em Stat Appl Genet Mol Biol}, 20(2):37--49, 07 2021.

\bibitem{Krishnanathan:2012tb}
Kirubhakaran Krishnanathan, Sean~R Anderson, Stephen~A Billings, and Visakan
  Kadirkamanathan.
\newblock A data-driven framework for identifying nonlinear dynamic models of
  genetic parts.
\newblock {\em ACS Synth Biol}, 1(8):375--84, Aug 2012.

\bibitem{Aquino:2014wt}
Gerardo Aquino, Luke Tweedy, Doris Heinrich, and Robert~G Endres.
\newblock Memory improves precision of cell sensing in fluctuating
  environments.
\newblock {\em Sci Rep}, 4:5688, Jul 2014.

\bibitem{Lakatos:2016hy}
Eszter Lakatos and Michael P~H Stumpf.
\newblock {Control mechanisms for stochastic biochemical systems via
  computation of reachable sets}.
\newblock {\em Roy Soc Open Sci}, page 079723, October 2016.

\bibitem{Harrington:2012cr}
Heather~A Harrington, Kenneth~L Ho, Thomas~W Thorne, Michael P~H Stumpf, and
  Michael P~H Stumpf.
\newblock {Parameter-free model discrimination criterion based on steady-state
  coplanarity.}
\newblock {\em Proc Natl Acad Sci USA}, 109(39):15746--15751, 2012.

\bibitem{Rashevsky1954}
N.~Rashevsky.
\newblock Topology and life: In search of general mathematical principles in
  biology and sociology.
\newblock {\em The Bulletin of Mathematical Biophysics}, 16:317--348, 1954.

\bibitem{Rosen1958}
Robert Rosen.
\newblock A relational theory of biological systems.
\newblock {\em The Bulletin of Mathematical Biophysics}, 20:245--260, 1958.

\bibitem{Eigen1971}
Manfred Eigen.
\newblock Selforganization of matter and the evolution of biological
  macromolecules.
\newblock {\em Die Naturwissenschaften}, 58:465--523, 1971.

\bibitem{Ganti2003}
Tibor G{\'{a}}nti.
\newblock {\em {The Principles of Life}}.
\newblock Oxford University Press, Oxford, UK, 2003.

\bibitem{Cornish_Bowden2011}
Athel Cornish-Bowden.
\newblock How far has it come?: Systems biology.
\newblock {\em The Biochemist}, 33:16--18, 2011.

\bibitem{Cleland2013}
Carol~E. Cleland.
\newblock Is a general theory of life possible? seeking the nature of life in
  the context of a single example.
\newblock {\em Biological Theory}, 7:368--379, 2013.

\bibitem{PWAnderson2011}
Philip~W. Anderson.
\newblock {\em {More and Different: notes from a thoughtful curmudgeon}}.
\newblock World Scientific, New Jersey, 2011.

\bibitem{Ryan2007}
Alex~J. Ryan.
\newblock Emergence is coupled to scope, not level.
\newblock {\em Complexity}, 13:67--77, 2007.

\bibitem{IrunRCohen2007}
Irun~R. Cohen and David Harel.
\newblock Explaining a complex living system: dynamics, multi-scaling and
  emergence.
\newblock {\em Journal of The Royal Society Interface}, 4:175--182, 2007.

\bibitem{Bedau2008}
Mark~A. Bedau.
\newblock Is weak emergence just in the mind?
\newblock {\em Minds and Machines}, 18:443--459, 2008.

\bibitem{Clayton2006}
Philip Clayton and Paul Davies, editors.
\newblock {\em {The Re-Emergence of Emergence: The Emergentist Hypothesis from
  Science to Religion}}.
\newblock Oxford University Press, New York, 2006.

\bibitem{Mayr1982}
Ernst Mayr.
\newblock {\em {The Growth of Biological Thought: Diversity, Evolution, and
  Inheritance}}.
\newblock Belknap Press, Cambridge, MA, 1982.

\bibitem{Van_Regenmortel2004}
Marc H.~V. {Van Regenmortel}.
\newblock Reductionism and complexity in molecular biology.
\newblock {\em {EMBO} Reports}, 5:1016--1020, 2004.

\bibitem{Bar_Yam2004}
Yaneer Bar-Yam.
\newblock A mathematical theory of strong emergence using multiscale variety.
\newblock {\em Complexity}, 9:15--24, 2004.

\bibitem{Thurston1994}
William~P. Thurston.
\newblock On proof and progress in mathematics.
\newblock {\em Bulletin of the American Mathematical Society}, 30:161--177,
  1994.

\bibitem{Cull2007}
Paul Cull.
\newblock The mathematical biophysics of {N}icolas {R}ashevsky.
\newblock {\em Biosystems}, 88:178--184, 2007.

\bibitem{Reed2015}
Michael~C. Reed.
\newblock Mathematical biology is good for mathematics.
\newblock {\em Notices of the American Mathematical Society}, 62:1172--1176,
  2015.

\bibitem{Borovik2021}
Alexandre Borovik.
\newblock A mathematician's view of the unreasonable ineffectiveness of
  mathematics in biology.
\newblock {\em Biosystems}, 205:104410, 2021.

\bibitem{Clairambault2013}
Jean Clairambault.
\newblock Can theorems help treat cancer?
\newblock {\em Journal of Mathematical Biology}, 66:1555--1558, 2013.

\bibitem{Scheiner2010}
Samuel~M. Scheiner.
\newblock Toward a conceptual framework for biology.
\newblock {\em The Quarterly Review of Biology}, 85:293--318, 2010.

\bibitem{Blagosklonny2002}
Mikhail~V. Blagosklonny and Arthur~B. Pardee.
\newblock Conceptual biology: Unearthing the gems.
\newblock {\em Nature}, 416:373, 2002.

\bibitem{Barnes2002}
Julie~C. Barnes.
\newblock Conceptual biology: a semantic issue and more.
\newblock {\em Nature}, 417:587--588, 2002.

\bibitem{Brenner2002}
Sydney Brenner.
\newblock Nature's gift to science --- {N}obel lecture, december 8.
\newblock https://www.nobelprize.org/prizes/medicine/2002/brenner/lecture/,
  2002.

\bibitem{Nurse2021}
Nurse.
\newblock Biology must generate ideas as well as data.
\newblock {\em Nature}, 597:305, 2021.

\bibitem{Noble2011}
Denis Noble.
\newblock Differential and integral views of genetics in computational systems
  biology.
\newblock {\em Interface Focus}, 1:7--15, 2011.

\bibitem{Noble2012}
Denis Noble.
\newblock A theory of biological relativity: no privileged level of causation.
\newblock {\em Interface Focus}, 2:55--64, 2012.

\bibitem{Rajapakse2017}
Indika Rajapakse and Stephen Smale.
\newblock Emergence of function from coordinated cells in a tissue.
\newblock {\em Proceedings of the National Academy of Sciences},
  114:1462--1467, 2017.

\bibitem{Sapoval:2022wo}
Nicolae Sapoval, Amirali Aghazadeh, Michael~G Nute, Dinler~A Antunes, Advait
  Balaji, Richard Baraniuk, C~J Barberan, Ruth Dannenfelser, Chen Dun,
  Mohammadamin Edrisi, R~A~Leo Elworth, Bryce Kille, Anastasios Kyrillidis,
  Luay Nakhleh, Cameron~R Wolfe, Zhi Yan, Vicky Yao, and Todd~J Treangen.
\newblock Current progress and open challenges for applying deep learning
  across the biosciences.
\newblock {\em Nat Commun}, 13(1):1728, 04 2022.

\bibitem{AlQuraishi:2021wr}
Mohammed AlQuraishi and Peter~K Sorger.
\newblock Differentiable biology: using deep learning for biophysics-based and
  data-driven modeling of molecular mechanisms.
\newblock {\em Nat Methods}, 18(10):1169--1180, 10 2021.

\bibitem{Naert:2021wc}
Thomas Naert, {\"O}zg{\"u}n {\c C}i{\c c}ek, Paulina Ogar, Max B{\"u}rgi,
  Nikko-Ideen Shaidani, Michael~M Kaminski, Yuxiao Xu, Kelli Grand, Marko
  Vujanovic, Daniel Prata, Friedhelm Hildebrandt, Thomas Brox, Olaf
  Ronneberger, Fabian~F Voigt, Fritjof Helmchen, Johannes Loffing, Marko~E
  Horb, Helen~Rankin Willsey, and Soeren~S Lienkamp.
\newblock Deep learning is widely applicable to phenotyping embryonic
  development and disease.
\newblock {\em Development}, 148(21), 11 2021.

\end{thebibliography}
\end{document}